# Cross-Section-Based Scaling Method for Material-Specific Cluster Dose Calculations – A Proof of Concept

Miriam Schwarze[1,2,*], Hui Khee Looe[2], Björn Poppe[2], Leo Thomas[1] and Hans Rabus[1]

[1] Physikalisch-Technische Bundesanstalt, Berlin, Germany
[2] Carl von Ossietzky Universität Oldenburg, Oldenburg, Germany

* Corresponding author, miriam.schwarze@ptb.de

**Abstract**

Cross-section data unavailability for non-water materials in track structure simulation software necessitates nanodosimetric quantity transformation from water to other materials. Cluster dose calculation transformation initially employed mass-density-based scaling - an approach resulting in a physically unrealistic material-independence of the cluster dose equation. This study introduces an alternative scaling method based on material-specific ionization cross-sections. The mean free path ratio of the materials for both the primary particles of the track structure simulation and for the secondary electrons served as the scaling factor. The approach was demonstrated through a cluster dose calculation for a carbon ion beam in a realistic head geometry and compared to the previous scaling method. The proposed cross-section-based scaling method resulted in a physically expected increase in cluster dose values for denser materials, which was not visible in the original scaling approach. The introduced scaling approach can be used to determine cluster dose distributions in heterogeneous geometries, a fundamental requirement for its integration into radiotherapy treatment planning frameworks.

## 1. Introduction

Nanodosimetry concepts provide a direct physics-based methodology for determining the biological effectiveness of ionizing radiation. The approach characterizes radiation quality via the spatial distribution of energy imparted by ionizations along the particle track within volumes comparable in size to DNA segments (typically 10 base pairs [1]). This distribution correlates with the biological effects of radiation [2]. Nanodosimetry thus offers an alternative approach for modeling the biological effects of radiation compared to the relative biological effectiveness (RBE) models currently used in carbon ion therapy treatment planning.

A connection between the nanodosimetric quantities and the macroscopic volumes used in radiation treatment planning (millimeter scale) is established by the cluster dose concept, introduced 2023 by Faddegon et al. [3]. For this purpose, the nanodosimetric quantities are weighted by the fluence of all particle types and energies within a voxel, summed up, and divided by the mass of the voxel $m_j$.

The quantities are typically obtained from Monte Carlo simulations. The fluence is determined using a condensed history (CH) simulation of the beam within the patient volume, where the cumulative track length of the primary particle and all secondary particles is calculated for discrete energy intervals. The nanodosimetric quantities for these particle classes are derived from a separate track structure (TS) simulation within a smaller volume (micrometer scale). While a CH simulation combines several interactions into a single step, a TS simulation explicitly simulates each individual interaction. This requires more detailed cross-section data, which are currently only available for a very small number of materials for the most relevant particles in the pertinent simulation tools.

As a result, the TS simulation for determining the nanodosimetric quantities is conducted in water (which is similar to human tissue) rather than in the actual voxel material. To account for this in the cluster dose calculation, Faddegon et al. [3] proposed scaling the cumulative track length by the ratio of the mass density of the voxel material $\rho_j$ to that of water $\rho_0$. This is based on the assumption that ionization density is proportional to material density. By scaling with the density ratio $\rho_j/\rho_0$, the material-dependent terms $\rho_j$ and $m_j$ combine to form the voxel volume $V_j = m_j/\rho_j$, which is constant. As a result, these terms effectively drop out of the cluster dose expression.

Since the nanodosimetric quantity is calculated in water and the cumulative track length remains unchanged across different materials, this leads to a cluster dose expression that no longer reflects any material dependence, which appears physically not meaningful.

In this work, we propose an alternative scaling method based on ionization cross-sections, inspired by the scaling procedure originally proposed by Grosswendt et al. for experimental nanodosimetry [1,4], which has been validated by Hilgers et al. [5]. The aim of the original scaling approach was to transform the mean number of ionizations measured in a gas-filled detection volume into a water-equivalent quantity. The scaling is based on the proportionality between the mean number of ionizations in a target and the ratio of the target diameter to the particle's mean free path for ionization. The methodology was tested using simulated cluster dose distributions for a carbon ion beam in a patient head geometry and compared to the original scaling approach.

## 2. Materials and Methods

*2.1 Nanodosimetric Quantities*

Ionizations occurring within a nanometer-sized target volume are grouped into clusters. The size of such a cluster is referred to as the Ionization Cluster Size (ICS) $v$. By determining the ICS in different target volumes along the particle track, a distribution of ICS values is obtained, which is called the Ionization Cluster Size Distribution (ICSD) $P(v)$. This distribution is the central quantity for characterizing radiation quality, and its derived quantities $I_P$, such as the cumulative frequency of clusters exceeding $k$ ionizations

$$F_k = \sum_{v=k}^{\infty} P(v), \qquad (1)$$

have been shown to correlate with DNA damage. In particular, higher values of $F_k$ (i.e. a higher probability of multiple ionizations within a nanometric volume) indicate the formation of dense ionization clusters. Such clusters are associated with complex DNA damage, including double-strand breaks, which are difficult to repair and therefore lead to a decreased cell survival probability [6].

*2.2 Cluster Dose Concept*

The cluster dose in voxel $j$ is defined as the sum over all particle classes $c$ of the number of ionization clusters produced by particles of this class in the voxel divided by the mass of the voxel $m_j$. The number of ionization clusters produced by particles of class $c$ is the product of the cumulative track length $t_j^c$ within voxel $j$ and the yield per length of the nanodosimetric quantity $I_P^c$ (i.e. $F_k$) produced by such a particle. Therefore, the cluster dose is described mathematically by the equation

$$g_j^{I_P} = \frac{1}{m_j} \sum_{c \in \mathcal{C}_j} t_j^c I_P^c. \qquad (2)$$

The particle classes $c$ distinguish between particle type and energy. The set of all classes is denoted by $\mathcal{C}$.

In the original formulation of the cluster dose concept [3], the cluster doses in voxel materials other than water are scaled according to the mass density $\rho$ in the voxel. The density scaling is independent of the particle class, leading to the following equation for the cluster dose

$$g_j^{I_P} = \frac{1}{m_j}\frac{\rho_j}{\rho_0} \sum_{c \in \mathcal{C}_j} t_j^c I_P^c = \frac{1}{\rho_0 V_j} \sum_{c \in \mathcal{C}_j} t_j^c I_P^c, \qquad (3)$$

where $V_j$ represents the voxel volume, $\rho_j$ the voxel density and $\rho_0$ the density of the material used in the TS simulation. Due to the applied density scaling, the equation becomes independent of the material in voxel $j$. The nanodosimetric quantity $I_P^c$ is calculated in water and is therefore material-independent in this equation. While the mean free path of the particle changes depending on the material, the sum of the step lengths $t_j^c$ remains the same for all materials.

*2.3 Scaling of Nanodosimetric Quantity*

The goal of scaling is to transform the ionization density distribution calculated in water, represented by the nanodosimetric quantity $I_p^c$, into a corresponding density distribution in another medium. This transformation depends on the ratio of ionization cross-sections between the target medium and water, both for the considered primary particle of class $c$ and for the secondary electrons produced by this particle, which also contribute to the ICSD.

Großwendt et al. [1,4] proposed a scaling approach to convert the mean ICS $M_1$, measured in gas, into a water-equivalent quantity. This approach assumes the mean number of ionizations within the target volume, $M_1$, is proportional to the ratio of the target volume's diameter $D$ to the mean free path $\lambda$ between successive ionizations by the primary particle multiplied by the mean number $\mu_1$ of ionizations in the target per ionizing interaction of the primary particle [1]

$$M_1 \propto \frac{D}{\lambda} \mu_1. \qquad (4)$$

The factor $\mu_1$ includes the contribution of secondary electrons and can be approximated by the mean ionization cross-section of electrons $\sigma^{e-}$ in the relevant energy range (assuming a material-independent electron energy spectrum). The mean free path is the inverse of the product of the cross-section $\sigma$ and the particle density $n = \rho/A$, where $A$ is the effective atomic mass and $\rho$ the material mass density of the material. Assuming equal-sized target volumes, this leads to the scaling factor

$$\frac{M_1^m}{M_1^w} = \frac{\sigma_m n_m}{\sigma_w n_w} \frac{\sigma_m^{e-}}{\sigma_w^{e-}} = \frac{\rho_m \sigma_m A_w}{\rho_w \sigma_w A_m} \frac{\sigma_m^{e-}}{\sigma_w^{e-}}. \qquad (5)$$

Since the cumulative frequencies $F_k$ are related to the mean ionization cluster size $M_1$ by a universal relation [7], this scaling relation can also be applied to $F_k$.

This results in the following expression for the cluster dose

$$g_j^{I_P} = \frac{1}{m_j} \sum_{c \in \mathcal{C}_j} \frac{\rho_j \sigma_j^c \sigma_j^{e-} A_0}{\rho_0 \sigma_0^c \sigma_0^{e-} A_j} t_j^c I_P^c = \frac{1}{\rho_0 V_j} \frac{A_0}{A_j} \sum_{c \in \mathcal{C}_j} \frac{\sigma_j^c \sigma_j^{e-}}{\sigma_0^c \sigma_0^{e-}} t_j^c I_P^c. \qquad (6)$$

with the cross-sections $\sigma^c$ of a particle of class $c$.

*2.4 Monte-Carlo Simulation*

The scaling effect on cluster dose distributions in an inhomogeneous volume was demonstrated using the realistic head geometry of a patient. The patient geometry corresponds to the *GLI_004_GBM* sample from the GLIS-RT dataset of the TCIA [8]. In this dataset, the geometry is provided as a CT image with a voxel size of (1.3 × 1.3 × 2.5) mm³. A circular carbon ion beam with a radius of 0.5 mm and varying initial energies and positions was used.

### 2.4.1 CH Simulation for Determining $t_j^c$

The CH simulation for determining the cumulative track lengths $t_j^c$ was performed using Geant4 [9–11]. The set $\mathcal{C}$ of particle classes includes the primary C-12 ion and the secondary particles C-11, C-10, B-11, B-10, protons, alpha particles, He-3, H-3, and H-2, along with energy bins ranging from 1 MeV to 3000 MeV. The bin width increases with energy [1]. The CT image was read using the Geant4 DICOM reader [12] and converted into a density distribution. Based on the density values, the materials were assigned to the voxels according to the ICRU material definitions [13]. Supplementary Table 1 lists the corresponding density intervals and material compositions. The ionization cross-sections were calculated using the *G4EmCalculator* for all particle classes and for electrons between 300 eV and 300 keV in all materials used in the simulation.

The physicslist used was derived from the *hadrontherapy* example [14]. The list includes the electromagnetic processes from the *G4EmStandardPhysics_option4* constructor, the decay processes from the *G4DecayPhysics* and *G4RadioactiveDecayPhysics* constructors, as well as the hadronic processes from the *G4IonBinaryCascadePhysics*, *G4EmExtraPhysics*, *G4HadronElasticPhysicsHP*, *G4StoppingPhysics*, *G4HadronPhysicsQGSP_BIC_HP*, and *G4NeutronTrackingCut* constructors. Using the *G4SteppingAction*, the step length for each step was determined and added to the cumulative track length for the respective particle class and voxel. A voxel boundary is treated as a volume boundary in Geant4, so a step is automatically executed when such a boundary is encountered.

### 2.4.2 TS Simulation for determining $I_p^c$

To determine the nanodosimetric quantity $I_p^c$, the TS simulation software Geant4-DNA [15–17] was used. For all particle classes, a monoenergetic point source with 50 000 primary particles was simulated. The volume comprised a cube with a side length of 150 μm made of *G4Water*. The source was positioned 0.5 μm behind the phantom edge in the beam direction inside the phantom and was centered in the lateral dimensions. The primary particle was stopped after 1 μm, and only secondary particles (electrons and photons) were transported thereafter. The electromagnetic processes from the *G4EmDNAPhysics_option4* constructor were used here. For each ionization, the location and identification number of the primary particle were recorded using the *G4SteppingAction*.

The water cube was placed within a cubic world volume of 1 cm edge length, also made of *G4Water*. In this region, the TS processes were disabled, and instead, the CH processes from the *G4EmStandardPhysics_option4* constructor were utilized.

---

[1] The centers of the energy bins are: [1, 2, 3, 4, 6, 10, 15, 20, 25, 30, 35, 41.25, 50, 60, 70, 80, 90, 103.75, 125, 150, 175, 200, 225, 250, 275, 300, 325, 350, 375, 400, 425, 450, 475, 500, 525, 550, 575, 600, 625, 650, 675, 700, 725, 750, 775, 800, 825, 850, 875, 900, 925, 950, 975, 1006.25, 1050, 1100, 1150, 1200, 1250, 1300, 1350, 1400, 1450, 1500, 1550, 1600, 1650, 1700, 1750, 1800, 1850, 1900, 1950, 2012.5, 2100, 2200, 2300, 2400, 2500, 2600, 2700, 2800, 2900, 3000] MeV. The bin boundaries were at the arithmetic mean of neighboring bin centers; the lower boundary of the first bin was 0.5 MeV and the higher boundary of the last at 3050 MeV.

The recorded ionizations were clustered using the fixed volume clustering method of Braunroth et al. [18] and the same target volume dimensions. The determined ICSDs were normalized to the number of primary particles and to the track length of the primary particle of 1 µm.

## 3. Results

Figure 1 and Figure 2 present a comparison of the cluster dose scaling for two different combinations of initial carbon ion beam energy and beam position (corresponding to a different material composition along the beam) for four different nanodosimetric quantities in the respective top rows. In both cases, the beam traverses the skull bone initially, then the brain tissue, and finally the skull bone once more. The dashed turquoise line represents the cluster dose profile along the beam axis with the density scaling according to Faddegon et al. [3] calculated with equation (3), while the solid yellow line indicates the cluster dose profile with the cross-section-based scaling calculated with equation (6). The mass density distribution along the beam axis is depicted with a black dashed line. The bottom row provides a comparison between the cluster dose profiles with cross-section-based scaling according to equation (6) and the depth dose curve. Again, the cluster dose profiles are plotted in a solid yellow line, while the dose profile is shown as a dashed red line. From left to right in each row, the panels correspond to the nanodosimetric quantities $F_2$, $F_4$, $F_5$, and $F_7$, calculated according to equation (1). Figure 1 displays the data for a 2000 MeV (166.66 MeV/u) carbon ion beam, whereas Figure 2 corresponds to a beam energy of 2750 MeV (229.16 MeV/u).

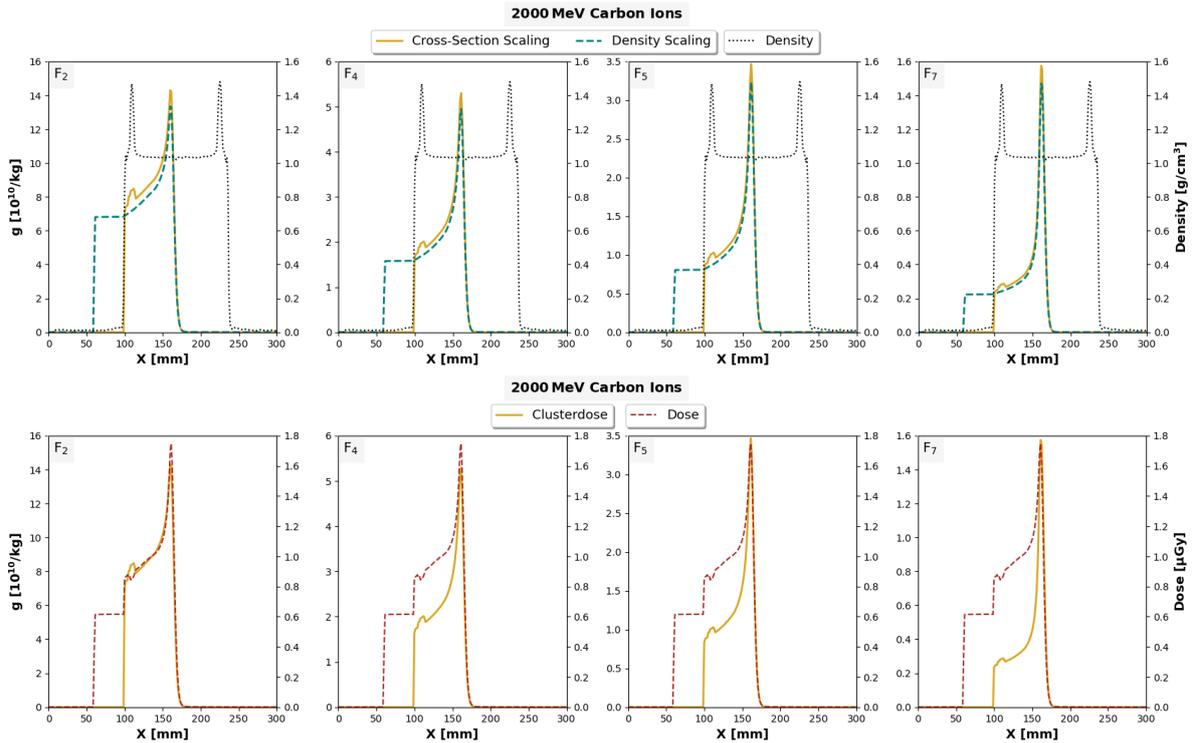

Figure 1: Top: Comparison of the cluster dose profiles along the beam axis with the density scaling of Faddegon et al. [2] according to equation (3) (dashed turquoise line) and the cross-section-based scaling according to equation (5) (solid yellow line). In addition, the mass density of the geometry along the beam axis is shown in the black dotted line. Bottom: Comparison of the cluster dose profiles with the cross-section-based scaling according to equation (5) (solid yellow line) and dose profiles (dashed red line) along the beam axis. Both rows show the data of a carbon ion beam with an initial energy of 2000 MeV (energy per mass of 166.66 MeV/u), propagating from left to right. The panels show the cluster dose profiles for different nanodosimetric quantities, from left to right: $F_2$, $F_4$, $F_5$, and $F_7$. All quantities are given per primary particle.

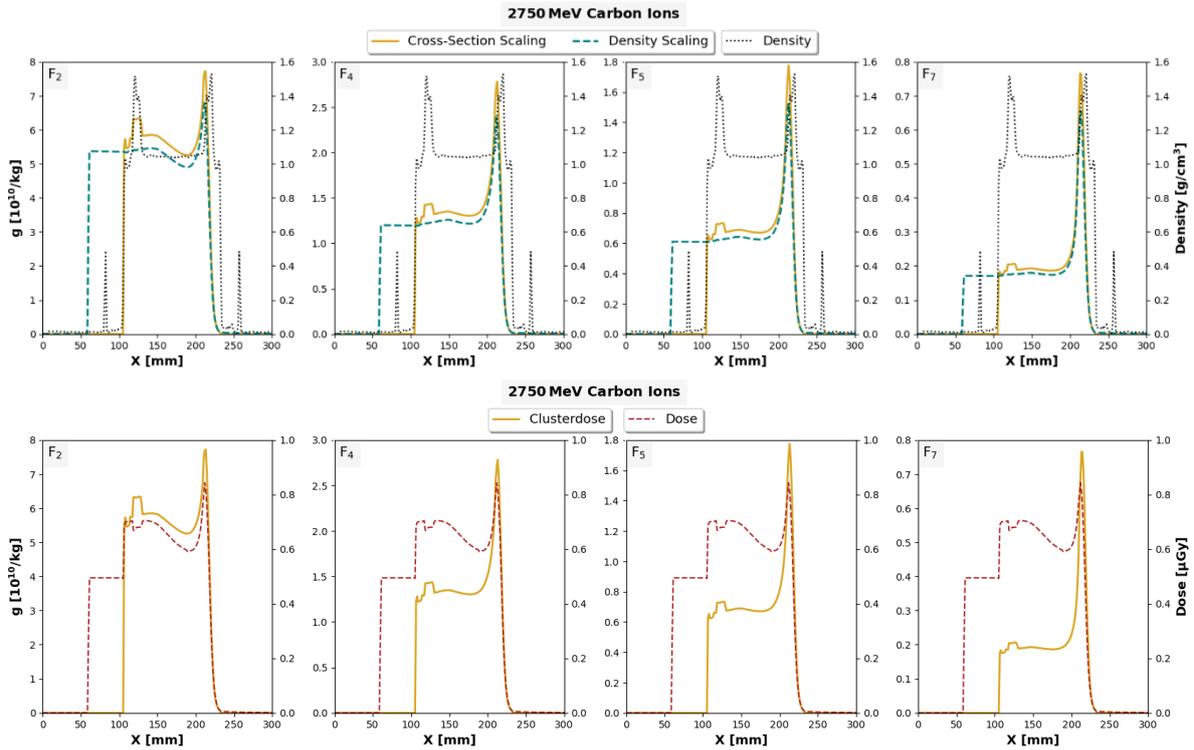

Figure 2: Top: Comparison of the cluster dose profiles along the beam axis with the density scaling of Faddegon et al. [3] according to equation (3) (dashed turquoise line) and the cross-section-based scaling according to equation (5) (solid yellow line). In addition, the mass density of the geometry along the beam axis is shown in the black dotted line. Bottom: Comparison of the cluster dose profiles with the cross-section-based scaling according to equation (5) (solid yellow line) and dose profiles (dashed red line) along the beam axis. Both rows show the data of a carbon ion beam with an initial energy of 2750 MeV (energy per mass of 229.16 MeV/u), propagating from left to right. The panels show the cluster dose profiles for different nanodosimetric quantities, from left to right: $F_2$, $F_4$, $F_5$, and $F_7$. All quantities are given per primary particle.

### 4. Discussion

As expected, the cluster dose profiles with the density scaling of Faddegon et al. [3] showed no change with density. This behavior does not reflect that the probability of ionization is increased in higher-density materials due to the higher electron density. This leads to a higher density of ionizations, a higher ICS and thus also a higher cluster dose. The new cross-section-based scaling cluster dose profiles reflect this effect. They do not show a high cluster dose in the air outside the body where owing to the lower density, no nanometric ionization clusters are expected. At present, validation of this scaling approach with TS simulations is not possible, as the required cross-section data are unavailable in the simulation codes.

It may be argued that the material composition near or in the cell nucleus can be expected to not depend much on the type of cell and that it is ionization clusters in DNA and the surrounding water shell that matter for the biological effectiveness [19]. From this position, the influence of the material composition on the cluster dose would be reflected in the fluence spectra as these are determined by the higher number of interactions in the denser medium.

However, aforementioned argument appears rather an objection to the cluster dose concept than to the remedy proposed here. By using voxel averages, the simulations leading to the data for calculation of the cluster dose use an effective medium instead of considering the changes of density and interactions on the microscopic scale. Therefore, it appears consequential to also consider the same effective medium in the scaling of the cluster dose from a simulation in water.

Which of the two approaches yields more realistic information on the ionization clusters that relate to biological radiation effects cannot be judged at present since deciding this would require tools for performing track structure simulation in media other than water.

Compared to the dose, the cluster dose profiles exhibit a more pronounced Bragg peak, particularly for $F_k$ with higher values of $k$. This is evident from the larger peak-to-plateau ratio in the cluster dose profiles and is due to the increased LET near the end of the particle range, which leads to a higher ionization density. Both profiles follow a similar overall trend.

In low-density regions such as the air surrounding the patient, the cluster dose decreases by several orders of magnitude to nearly zero, whereas the dose profile shows only a moderate reduction compared to tissue. The steep decline in the cluster dose values in air can be attributed to the low probability of ionization events. As a result, only isolated ionizations occur within the nanometer-sized target volumes. The dose concept does not account for this spatial distribution; instead, it considers the energy deposited per unit mass, averaged over a macroscopic volume. Even if energy is deposited sparsely across large distances, the resulting energy per mass can still be significant, leading to a nonzero dose.

Notably, voxel-based scaling can introduce inaccuracies at interfaces with steep density gradients. The increased production of secondary electrons in the denser material leads to a locally elevated number of ionizations near such interfaces. A voxel-based scaling approach excludes this effect. However, since secondary electrons have a limited range of only a few micrometers, the resulting inaccuracies are considered minor.

Furthermore, the presented scaling method is limited to nanodosimetric quantities for which a linear relationship with interaction probability exists. This condition is fulfilled for quantities that describe the distribution core. Toward the tails of the ICSD, geometrical factors such as the target volume size become increasingly relevant. These factors remain unaccounted for by both scaling methods.

For this proof-of-principle demonstration of the cross-section-based scaling method, the mean ionization cross-section of electrons within the relevant energy range was used for each material, as the cross-sections exhibit no significant variation across this range. However, other application scenarios may require electron energy spectrum determination and ionization cross-section weighted averaging.

In addition, our study was limited to a simple pencil beam. However, in the clinical context, superpositions of multiple pencil beams are applied, resulting in a broadened Bragg peak known as the spread-out Bragg peak. Cluster dose calculations for such radiation fields have already been tested in a homogeneous medium [20] and recently in a realistic, heterogeneous geometry with the original density scaling [21] and should also be tested for the cross-section-based scaling method in future studies. While different radiation field configurations would primarily affect the CH simulation and lead to a modified distribution of cumulative track length, the underlying nanodosimetric quantities and their material dependence are not influenced by the beam composition itself. Therefore, the scaling approach should remain applicable to such irradiation scenarios.

## 5. Conclusions

To apply the cluster dose framework in treatment planning, calculating the cluster dose in inhomogeneous geometries is necessary. The previously used equation for cluster dose calculation [3] exhibits material independence, which arises from scaling the nanodosimetric quantity using the ratio of mass densities. Our proposed alternative scaling approach, based on material-specific ionization cross-sections, overcomes this limitation by enabling the cluster dose determination in patient geometries.

## 6. Acknowledgments

This work was in part supported by the "Metrology for Artificial Intelligence in Medicine (M4AIM)" program, funded by the German Federal Ministry of Economic Affairs and Climate Action in the frame of the QI-Digital Initiative.

## 7. Data Availability Statement

The data that support the findings of this study are openly available at the following URL: https://gitlab1.ptb.de/MiriamSchwarze/clusterdosesimulation.

**Supplement**

Supplementary Table 1: Composition and density intervals (the respective upper bounds of the intervals are given) of the materials used in the Geant4 condensed-history simulation.

| Material | Upper Density Bound [g/cm³] | Composition | |
|---|---|---|---|
| | | Fraction [%] | Element |
| Air | 0.5156 | 70 | N |
| | | 30 | O |
| SoftTissue | 1.035 | 10.4472 | H |
| | | 23.219 | C |
| | | 2.488 | N |
| | | 63.0238 | O |
| | | 0.113 | Na |
| | | 0.0113 | Mg |
| | | 0.113 | P |
| | | 0.199 | S |
| | | 0.134 | Cl |
| | | 0.199 | K |
| | | 0.023 | Ca |
| | | 0.005 | Fe |
| | | 0.003 | Zn |
| Brain | 1.07 | 11.0667 | H |
| | | 12.542 | C |
| | | 1.328 | N |
| | | 73.7723 | O |
| | | 0.1840 | Na |
| | | 0.015 | Mg |
| | | 0.356 | P |
| | | 0.177 | S |
| | | 0.236 | Cl |
| | | 0.31 | K |
| | | 0.009 | Ca |
| | | 0.005 | Fe |
| | | 0.001 | Zn |
| SpinalDisc | 1.14 | 9.60 | H |
| | | 9.90 | C |
| | | 2.20 | N |
| | | 74.40 | O |
| | | 0.50 | Na |
| | | 2.20 | P |
| | | 0.90 | S |
| | | 0.30 | Cl |
| TrabecularBone | 1.55 | 8.50 | H |
| | | 40.40 | C |
| | | 2.80 | N |
| | | 36.70 | O |

| | | 0.10 | Na |
| | | 0.10 | Mg |
| | | 3.40 | P |
| | | 0.20 | S |
| | | 0.20 | Cl |
| | | 0.10 | K |
| | | 7.40 | Ca |
| | | 0.10 | Fe |
| CorticalBone | 2.03 | 4.7234 | H |
| | | 14.4330 | C |
| | | 4.199 | N |
| | | 44.6096 | O |
| | | 0.22 | Mg |
| | | 10.497 | P |
| | | 0.315 | S |
| | | 20.993 | Ca |
| | | 0.01 | Zn |
| ToothDentine | 2.515 | 2.67 | H |
| | | 12.77 | C |
| | | 4.27 | N |
| | | 40.40 | O |
| | | 0.65 | Na |
| | | 0.59 | Mg |
| | | 11.86 | P |
| | | 0.04 | Cl |
| | | 26.74 | Ca |
| | | 0.01 | Zn |
| ToothEnamel | 3.265 | 0.95 | H |
| | | 1.11 | C |
| | | 0.23 | N |
| | | 41.66 | O |
| | | 0.79 | Na |
| | | 0.23 | Mg |
| | | 18.71 | P |
| | | 0.34 | Cl |
| | | 35.97 | Ca |
| | | 0.02 | Zn |